\documentclass[letterpaper, 10 pt, conference]{ieeeconf}   

\IEEEoverridecommandlockouts                              
\usepackage{mathtools}
\usepackage{enumerate}
\usepackage{amsmath}
\usepackage{amssymb}
\usepackage{hyperref}
\usepackage{graphicx}
\usepackage{booktabs}
\usepackage[english]{babel}
\usepackage{xcolor, soul}
\usepackage{bm}
\usepackage{etoolbox} 
\usepackage{verbatim}
\usepackage{pgf}
\usepackage{algorithm}
\usepackage{algpseudocode}
\usepackage{gensymb}

\newcommand{\spaceOpti}{\hspace{-0.5cm}}
\hypersetup{
    colorlinks=false,
    pdfborder={0 0 0},
}
\def\pgfscale{0.78} 

\title{A Learning-based Model Predictive Control Scheme with Application to Temperature Control Units}

\author{Jing Xie$^{1, *}$, Léo Simpson $^{2, *}$\thanks{$^*$ Equal contribution}, Jonas Asprion$^{2}$, Riccardo Scattolini$^{1}$
	\thanks{$^{1}$ The authors are with Dipartimento di Elettronica, Informazione e Bioingegneria, Politecnico di Milano, Via Ponzio 34/5, 20133, Milano, Italy.}
 \thanks{$^{2}$ The authors are with Tool-Temp AG, Industriestrasse 30, 8583, Sulgen, Switzerland. }}

\begin{document}

\maketitle
   			
\begin{abstract}
Temperature control is a complex task due to its often unknown dynamics and disturbances. This paper explores the use of Neural Nonlinear AutoRegressive eXogenous (NNARX) models for nonlinear system identification and model predictive control of a temperature control unit. First, the NNARX model is identified from input-output data collected from the real plant, and a state-space representation with known measurable states consisting of past input and output variables is formulated. Second, a tailored model predictive controller is designed based on the trained NNARX network.
The proposed control architecture is experimentally tested on the temperature control units manufactured by Tool-Temp AG. The results achieved are compared with those obtained using a PI controller and a linear MPC. The findings illustrate that the proposed scheme achieves satisfactory tracking performance while incurring the lowest energy cost among the compared controllers.
\end{abstract}

\begin{keywords}
	Nonlinear Model Predictive Control, Neural Networks, Temperature Control
\end{keywords}

\section{Introduction}
\begin{figure}[t]
        \vspace{0.3cm}
	\centering
	\includegraphics[height=5cm]{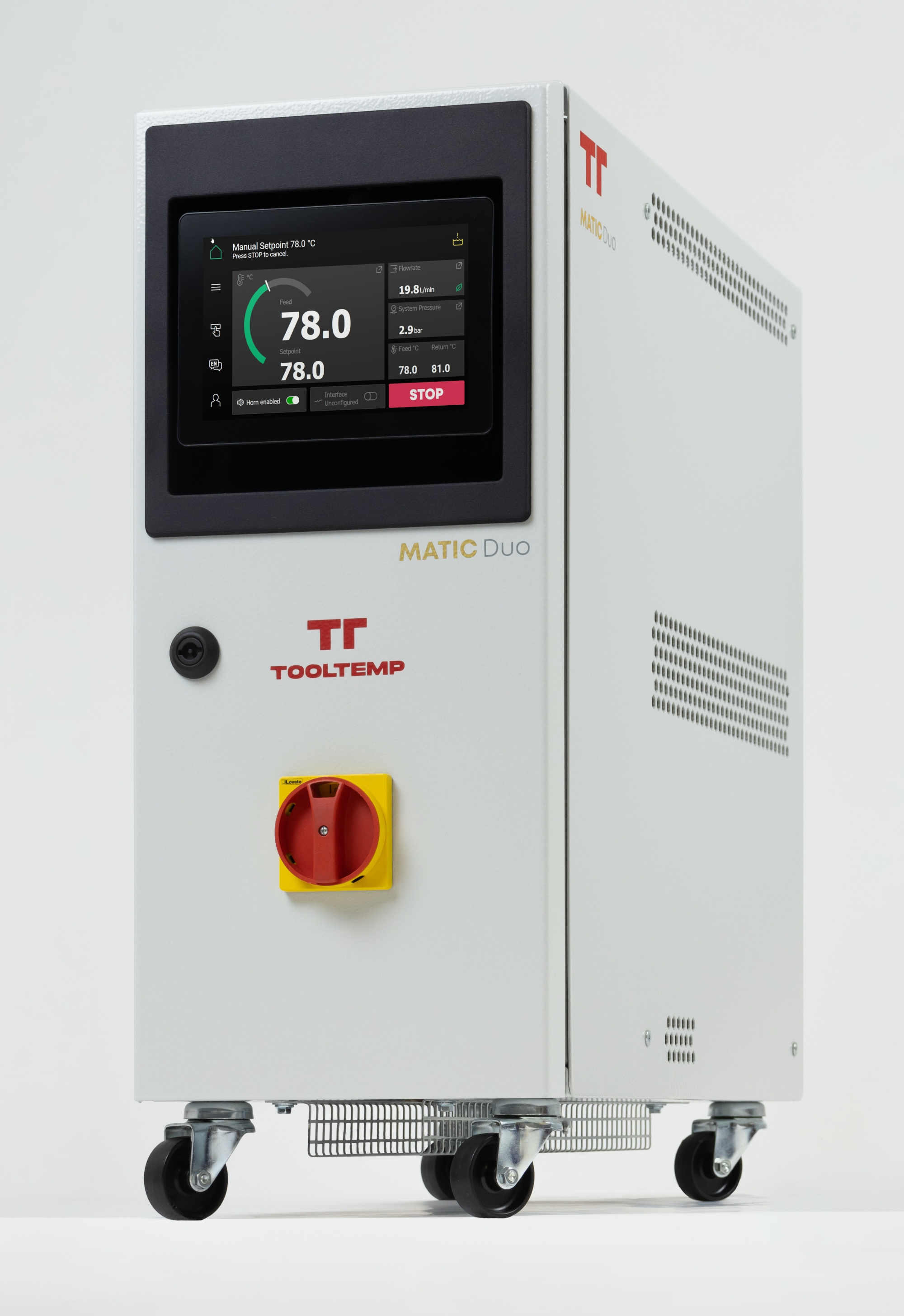}
	\caption{Temperature Control Unit manufactured by Tool-Temp AG}
	\label{fig:tcu}
\end{figure}

Temperature control is a critical aspect in a wide array of industrial, commercial, and residential applications, ranging from manufacturing processes \cite{cahyo2019co,rsetam2022robust} to climate control systems \cite{SOLANO2021269}. The ability to track specific temperature profiles is paramount for ensuring product quality or process efficiency. Temperature control units, also known as temperature controllers, play a pivotal role in achieving precise and stable temperature conditions within diverse operational environments.
To that end, temperature control units (TCUs), such as the ones produced by Tool-Temp AG, are used. These devices (see Figure \ref{fig:tcu}) temper the external process by supplying a thermal fluid at a suitable temperature and volumetric flow. Due to their universal applicability, TCUs are often used for different tasks on an hourly to daily basis. 

In practical applications, these devices are commonly governed by a PID controller, owing to its ease of implementation and remarkable performance \cite{McMillan2012}. Nevertheless, PID control exhibits limitations, such as the requirement for precise tuning, sensitivity to system changes, and challenges in handling nonlinear processes \cite{johnson2005pid}. While widely used for their simplicity, these limitations have led to the development and application of more advanced control strategies, e.g., model-based control. The adoption of model-based control strategies provides enhanced flexibility, precision, and adaptability in handling complex systems \cite{HOU20133}.
Indeed, accurate mathematical models are key to the implementation of such control algorithms.
The common modeling practice for TCUs heavily relies on physical knowledge, e.g., heat transfer behaviors for thermal systems.
Nevertheless, certain variables remain unknown to the TCU controller in the field.
These uncertainties include delays attributable to the tubes linking the TCU and external systems, the heat-transfer fluid used and its properties, as well as fluctuations in cooling water temperature and flow rate. All these factors contribute to the complexity of physical modeling, rendering it sensitive to unmodeled uncertainties.

While parametric linear systems could fit those requirements, enlarging the space of possible models with nonlinear ones might increase the modeling performance.
To this end, Recurrent Neural Networks (RNNs) emerge as a promising framework, owing to their robust ability to characterize nonlinear behaviors and make accurate time-series predictions. Many RNN model architectures are proposed in the literature, e.g., Long short-term memory networks (LSTM) \cite{hochreiter1997long}, Gated Recurrent Units (GRU) \cite{chung2014empirical}, Echo State Networks (ESN) \cite{jaeger2007echo} and Neural NARXs (NNARX) \cite{levin1996control}. 

Regarding model-based control algorithms, Model Predictive Control (MPC) is one of the most popular methods.
It uses process models to predict future system behaviors and minimize its cost.
At each time step, it solves a finite-dimensional constrained optimization problem, which approximates the optimality condition of the decision \cite{rawlings2017model}.

Given the latter discussions on model identification and model-based control, the goal of this paper is twofold. First, it focuses on the system identification of TCUs, including the identification of a linear model, and the training of Neural NARX networks. The choice of NNARX is due to its simple structure and ease of training. Second, a tailored MPC algorithm, leveraging the learned model, is designed for real-time temperature control of TCUs with tracking and economic objectives.
The paper is organized as follows. In Section \ref{sec:pro_state}, the physical system of TCU is described, and the control problem is formulated. In Section \ref{sec:sysiden} the training procedure is introduced. In Section \ref{sec:control}, a tailored nonlinear MPC algorithm is designed for real-time implementation. In Section \ref{sec:exp}, the experiments are conducted on TCUs, and performances of nonlinear MPC, linear MPC, and PI control are discussed. Lastly, some conclusions are drawn in Section \ref{sec:conclu}.

\section{Problem Statement}\label{sec:pro_state}

\begin{figure}[t]
	\centering	\includegraphics[width=\linewidth]{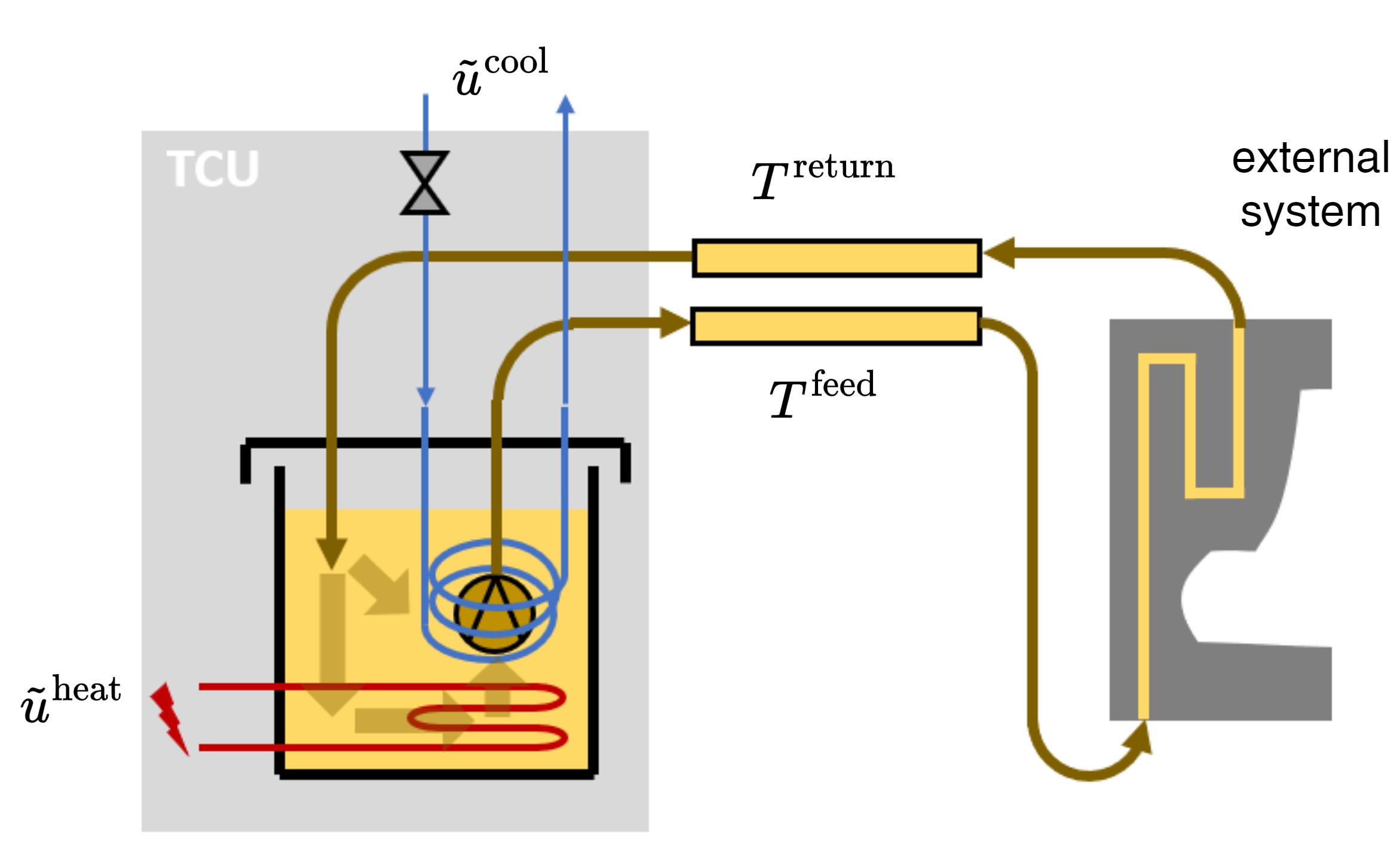}
	\caption{Overall system with TCU and external system}
	\label{fig:physics_system}
\end{figure}

The objective is to control the output fluid temperature $T^{\textup{feed}}$ of the TCU to match a predetermined reference temperature $r$. The TCU operates based on two binary control inputs.
The first, named $\tilde{u}^{\textup{heat}}$, represents heating by applying a voltage to electrical resistance, whereas 
the second, namely $\tilde{u}^{\textup{cool}}$ indicates the binary position of a valve that can let the cooling-water flow through a cooling coil.
The fluid, whether water or oil, is fed into an unknown external system at temperature $T^{\textup{feed}}$. After going through the unknown thermal process, it returns to the TCU at temperature $T^{\textup{return}}$.
These two latter temperatures are measured with sensors.
This is illustrated in Figure \ref{fig:physics_system}.

Every $\Delta t=1\textup{s}$, the binary control input $\tilde{u} = [\tilde{u}^{\textup{heat}}, \tilde{u}^{\textup{cool}}]$ is executed and $T^{\textup{feed}}$ is measured by sensors. 
\begin{figure}
	\vspace{0.3cm}
	\begin{center}
             \scalebox{\pgfscale}{
        	\input{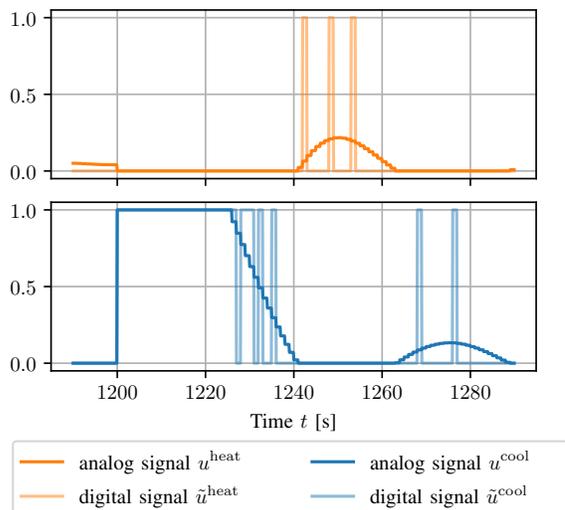}
            		}
	\end{center}
	\vspace{-0.3cm}
	\caption{Analog input signal $u$ together with its digital conversion $\tilde{u}$.}\label{fig:DeltaModulation}
\end{figure}
\begin{figure}[t]
	\centering
\includegraphics[width=0.9\linewidth]{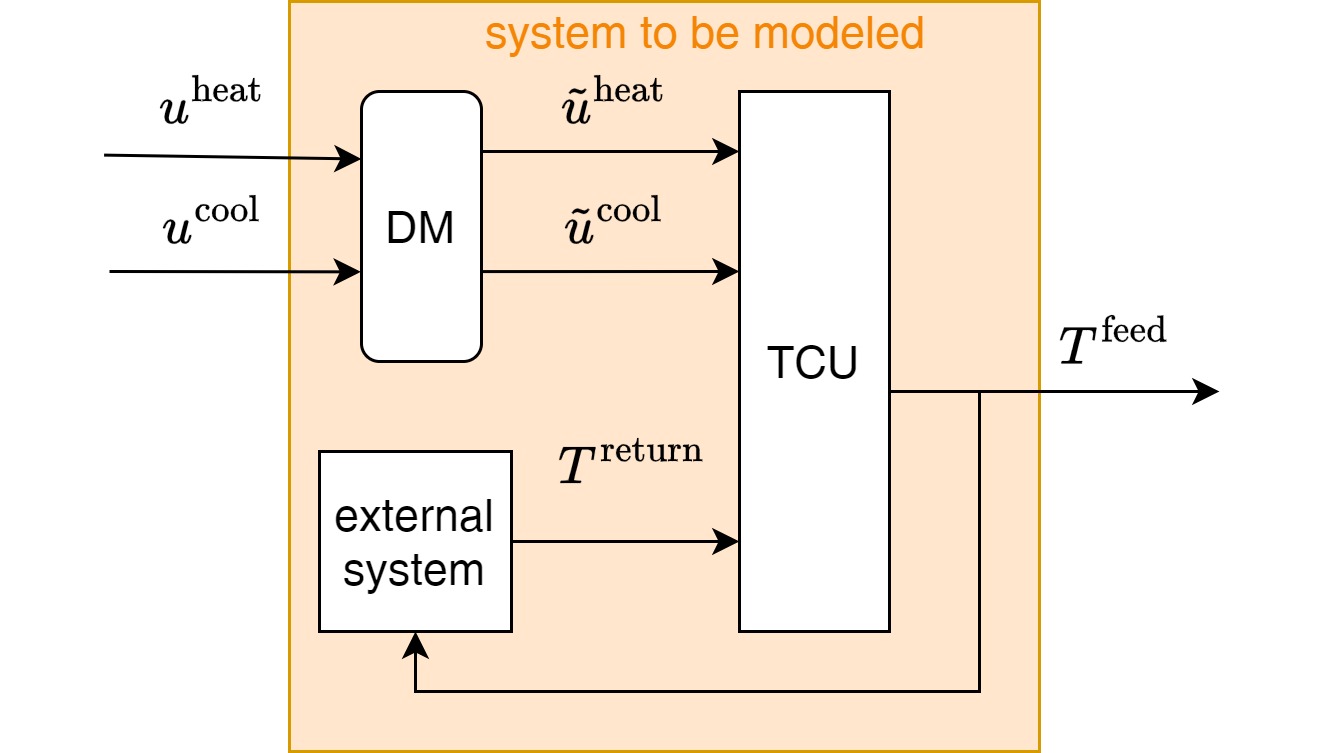}
	\caption{System abstraction}
	\label{fig:abstract_system}
\end{figure}
We denote $u = [u^{\textup{heat}},u^{\textup{cool}}]  \in [0,1]^2$ as the analog control input, which is computed by a discrete-time controller. At time step $k$, an analog input is transformed into a digital one using 
the delta modulation (DM) \cite{Schindler1970Delta} defined as follows:
\begin{align}\label{eq:modulation}
    &\tilde{u}^m_k \coloneqq
    \begin{cases}
        1 &\mbox{if} \; \sum\limits_{i=0}^{k-1}\tilde{u}^m_i < \sum\limits_{i=0}^{k}u^m_i    \\
        0 &\mbox{else}
    \end{cases}
\end{align}
where $m \in \left \{\textup{heat},\textup{cool} \right \}$.
An example signal of analog-to-digital conversion is depicted in Figure \ref{fig:DeltaModulation}.

The system that we choose to identify contains the TCU, the external process, and the analog-to-digital converter. Hence, it inputs the analog variables $u^{\textup{heat}}$ and $u^{\textup{cool}}$ and its single output is $T^{\textup{feed}}$.
This is depicted in Figure  \ref{fig:abstract_system}. 

In the experiments conducted for this paper, the external system is only a closed circuit, that acts as a delay for the temperature of the fluid. As a result, the system is mainly defined by the TCU.
However, in other applications of this device, the external system can be any other thermal process.

\section{System Identification} \label{sec:sysiden} 

In this section, we present the system identification procedure that has been conducted.
We first describe the experiment design that is used to collect data.
In the following parts, we use this data to identify two models: a linear model, and a black-box nonlinear model, using the NNARX structure.
The identified models will be used to design a Model Predictive Controller in Section \ref{sec:control}.

\subsection{Data Collection using a PI controller}\label{subsec:PI}

\begin{figure}
	\vspace{0.2cm}
	\begin{center}
             \scalebox{\pgfscale}{          			\input{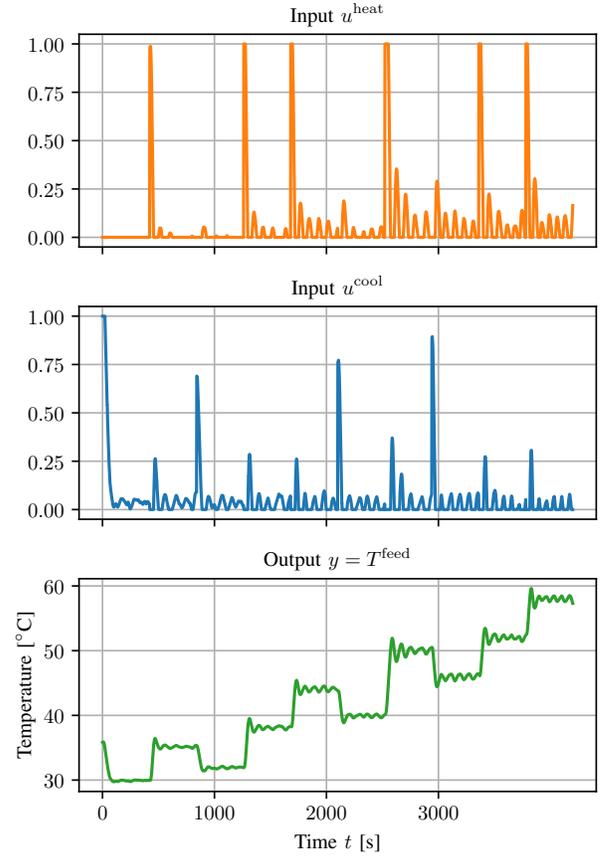} }
	\end{center}
	\vspace{-0.3cm}
	\caption{Exemplary sample of the collected data}\label{fig:data_collected}
\end{figure}

The first step of the model identification procedure is the data collection.
For this purpose, we conduct experiments on the real plant, while it is controlled by a simple control algorithm.
As opposed to generating inputs randomly, using closed-loop data is advantageous because it can guarantee safety during operation, and it allows to collect data for a chosen temperature range. Furthermore, the collected data will offer information that is relevant to our control task, such as steady-state information.
Note that in the case of closed-loop data, the data is informative as long as the reference signal that the controller tracks is itself persistently exciting \cite[Theorem 13.2]{ljung1999system}.
For convenience, the same controller will be used to assess the performance of our control algorithm in the last section of this paper.
For this simple controller, we choose a discrete Proportional-Integral (PI) Controller, defined by the following equation
\begin{equation}\label{eq:pi}
    u^{\textup{pi}}_k = K_pe_k + K_i\sum\limits_{i=0}^{k}e_i
\end{equation}
where $e_k \coloneqq r_k - y_k$ is the tracking error at time step $k$ and $y_k \eqqcolon T^{\textup{feed}}_k$ denotes the plant output at time step $k$. $K_i$ and $K_p$ denote parameters for the PI controller, which are empirically tuned to stabilize the plant. Eventually, $u^{\textup{pi}}$ is projected in the interval $[-1, 1]$ such that positive (negative) values indicate heating (cooling).
This is summarized by the following equations
\begin{align}
u^{\textup{heat}}=
     \begin{cases}
1,   &\mbox{if} \; u^{\textup{pi}} \geq 1 \\
0,  &\mbox{if} \; u^{\textup{pi}}\leq 0 \\
u^{\textup{pi}},   &\mbox{else}
     \end{cases}, \;
u^{\textup{cool}}=
     \begin{cases}
1 &\mbox{if} \; u^{\textup{pi}} \leq -1 \\
0, &\mbox{if} \; u^{\textup{pi}}\geq 0 \\
-u^{\textup{pi}},  &\mbox{else}
     \end{cases} 
\end{align}
An anti-windup strategy is also used, where the sum in \eqref{eq:pi} is not updated when the actuators are saturated. 
The output reference of PI is piecewise constant, and carefully chosen to generate an informative dataset.
The input-output data has been recorded with sampling time $\Delta t =1\textup{s}$ for a total of $11760$ time steps.
One of the collected sequences is shown in Figure \ref{fig:data_collected}. Finally, the data is resampled to $\tau_t = 6\textup{s}$ for model identification.

\subsection{Identification of a linear model}\label{subsec:lmpc}
Using the collected data, we estimate the parameters of a discrete-time state-space model that takes the following form
    \begin{align}
        x^1_{k+1} &= x^1_k + b_{\textup{h}}  u^{\textup{heat}}_k + b_{\textup{c}} u^{\textup{cool}}_k + n_1, \nonumber \\
        x^2_{k+1} &= x^2_k + a \big[ x^1_{k} - x^2_k \big]  + n_2,  \\
        T^{\textup{feed}}_k &= x^2_k + n_y, \nonumber
    \end{align}
where
$n_1, n_2, n_y$ are process and measurement noise,
$a, b_{\textup{h}}, b_{\textup{c}} $ are parameters of the model, and $x^1, x^2$ are the states of the system.
The model parameters together with the noise covariances are estimated from the data using the Maximum Likelihood Estimation (MLE) framework introduced in \cite{kashyap1970maximum}.
The algorithm that is used for the estimation procedure is described in \cite{simpson2023efficient}.
The identified nominal model can also be written compactly:
\begin{equation} \label{eq:linear_nominal}
	\Sigma_l: \begin{dcases}
		x_{k+1} = f_l(x_{k},u_{k}) \\
		y_k = g_l(x_k)
	\end{dcases}
\end{equation}
where $u = \big[ u^{\textup{heat}}, u^{\textup{cool}} \big]' $ are the input of the system, and $y = T^{\textup{feed}}$ is the output.

\subsection{Neural NARX models}

NNARX models \cite{bonassi2021nnarx} are nonlinear, time-invariant, discrete-time models.
 At time instant $k$, the future output $y_{k+1}$ is computed as a nonlinear regression function $\eta$ on past $N$ input $u_k \in {R}^m$ and output data $y_k \in {R}^p$, i.e.,
\begin{equation} \label{eq:rnn:nnarx_model}
y_{k+1} = \eta(y_{k}, y_{k-1}, ..., y_{k-N+1}, u_{k}, u_{k-1}, ... u_{k-N}; \bm{\Phi}),
\end{equation}
where $\bm{\Phi}$ is the training parameters.
By defining, for $i \in \{1, ..., N \}$,
\begin{equation} \label{eq:rnn:nnarx_states}
	z_{i, k} = \left[\begin{array}{c}
		y_{k-N+i} \\
		u_{k-N-1+i}
	\end{array}\right],
\end{equation}
and by denoting the state vector $x_{k} = [ z_{1, k}^\prime, ..., z_{N, k}^\prime]^\prime \in \mathbb{R}^{n}$, it is straightforward to rewrite model \eqref{eq:rnn:nnarx_model} in state space form:
\begin{equation} \label{eq:nnarx:statespace}
\begin{dcases}
  x_{k+1} = {A} x_{k} + B_{u} u_{k} + B_{x} \eta(x_{k}, u_{k}; \bm{\Phi}) \\
  y_{k} = C x_{k}
\end{dcases}
\end{equation}
where $A$, $B_u$, $B_x$, and $C$ are fixed matrices with known structure and elements equal to zero or one, see \cite{xie2023robustmpc}.
\begin{figure}[t]
	\centering
	\includegraphics[width=\linewidth]{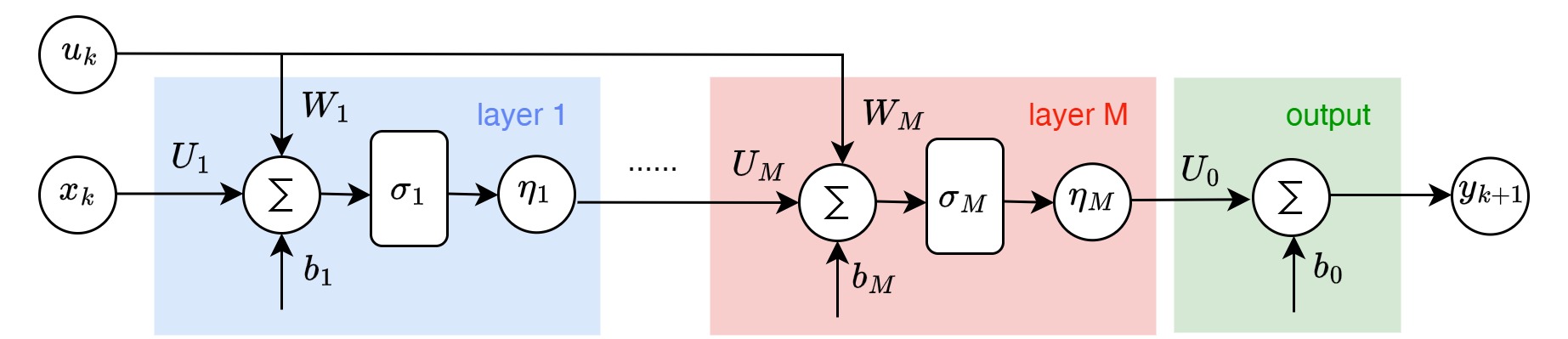}
	\caption{Structure of the NNARX model}
	\label{fig:nnarx}
\end{figure}
In NNARX models, the regression function $\eta$ in \eqref{eq:rnn:nnarx_model} is described by a Feed-Forward Neural Network (FFNN), which is depicted in Figure \ref{fig:nnarx}.
Each layer consists of a linear combination of its input and a bias, squashed by a suitable nonlinear function $\sigma$, known as the activation function. A compact formulation of $\eta$ is as follows
\begin{equation}  \label{eq:model:ffnn}
		\eta(x_{k}, u_{k}) = U_0\eta_M(\eta_{M-1}(...\eta_1(x_k,u_k),u_k),u_k)+b_0
\end{equation}
where $\eta_l$ is the nonlinear relation between $l$-th and the previous layer, which can be written as
\begin{equation}  \label{eq:model:ffnn2}
		\eta_l(\eta_{l-1},u_k) = \psi_l \big( W_l u_{k} + U_l \eta_{l-1} + b_l \big),
\end{equation}
$\psi_l$ is an activation function,  which is applied element-wise on its argument. We assume that $\psi_l$ is Lipschitz-continuous and satisfying $\psi_l(0) = 0$. The matrices $W_l$, $U_l$, and $b_l$ are the learnable weights of the layer, which constitute the network's training parameters
\begin{equation}
\Phi = \{ U_0, b_0, \{ U_l, W_l, b_l \}_{ l = 1, ..., M} \}
\end{equation}
Henceforth, the NNARX model is denoted as,
\begin{equation} \label{eq:nnarx:compact}
	\Sigma_n: \begin{dcases}
		x_{k+1} = f_n(x_{k},u_{k}) \\
		y_k = g_n(x_k)
	\end{dcases},
\end{equation}
where the dependency on $\bm{\Phi}$ is omitted for compactness. 

\subsection{NNARX model training}
The weights $\Phi$ as discussed are learned from the input-output data collected in Section \ref{subsec:PI}, where the weights that best explain the measured data are sought.
The activation function is $\tanh$ function in this paper, for more details, see \cite{bonassi2021nnarx} for example.

According to the Truncated Back-Propagation Through Time (TBPTT) principle \cite{bianchi2017recurrent}, $N_{t} = 84$ random subsequences of length $N_s = 133$ time steps, denoted by $(\bm{u}_{N_s}^{\{i\}}, \bm{y}_{p, N_s}^{\{i\}})$, with $i = 1, \dots , N_t$, have been extracted from the experimental data.
One other experiment has been performed, from which $N_v=32$ subsequences have been extracted as independent validation test datasets.

The training procedure has been conducted using PyTorch 1.9.
The employed NNARX model is characterized by a single-layer ($M=1$) FFNN regression function with $10$ neurons, and the past $N=8$ input-output data has been stored in the state.
Following the guidelines of \cite{bonassi2022survey, bonassi2021nnarx}, the model has been trained by minimizing the Mean Square Error (MSE) of one-step prediction over the training set, defined as follows
\begin{equation}\label{eq:criterion}
\mathcal{L}_{\boldsymbol{\text{MSE}}}(\bm{\Phi}) = \alpha
\sum_{i=1}^{N_t} \sum_{k=N_{w}}^{N_s} \Big( y_{k+1}(x_k^{\{ i \}},u_k^{\{ i\}})- y_k^{\{i\}} \Big)^2
\end{equation}
where $y_{k+1}(x_k^{\{ i\}},u_k^{\{ i\}})$ is the one-step prediction defined in \eqref{eq:rnn:nnarx_model}, $y_k^{\{i\}}$ the measurement value and $\alpha \coloneqq \frac{1}{N_t(N_s-N_{w})}$ is for scaling purposes. $N_w$ denotes the washout length to discount initialization. 
The NNARX model has been trained until the criterion function \eqref{eq:criterion} evaluated on the validation set becomes lower than a tolerance, set to $10^{-4}$.
In the present case, this corresponds to about $1500$ epochs.

The open-loop predictions are depicted in Figure \ref{fig:predictions} to compare the modeling performance of the linear model and NNARX model. The predictions start at four different points with prediction horizon $N_T=70$ steps.
The NNARX model provides better predictions than the linear one due to the past information stored in the state.
\begin{figure}
	\vspace{0.3cm}
	\begin{center}
             \scalebox{\pgfscale}{
        	\input{Figures/experiment_both_predictions_3.pgf} 
            		}
	\end{center}
	\vspace{-0.3cm}
	\caption{Open-loop prediction performance}\label{fig:predictions}
\end{figure}

\section{Control Scheme}\label{sec:control}

\begin{figure}[t]
	\centering	\includegraphics[width=\linewidth]{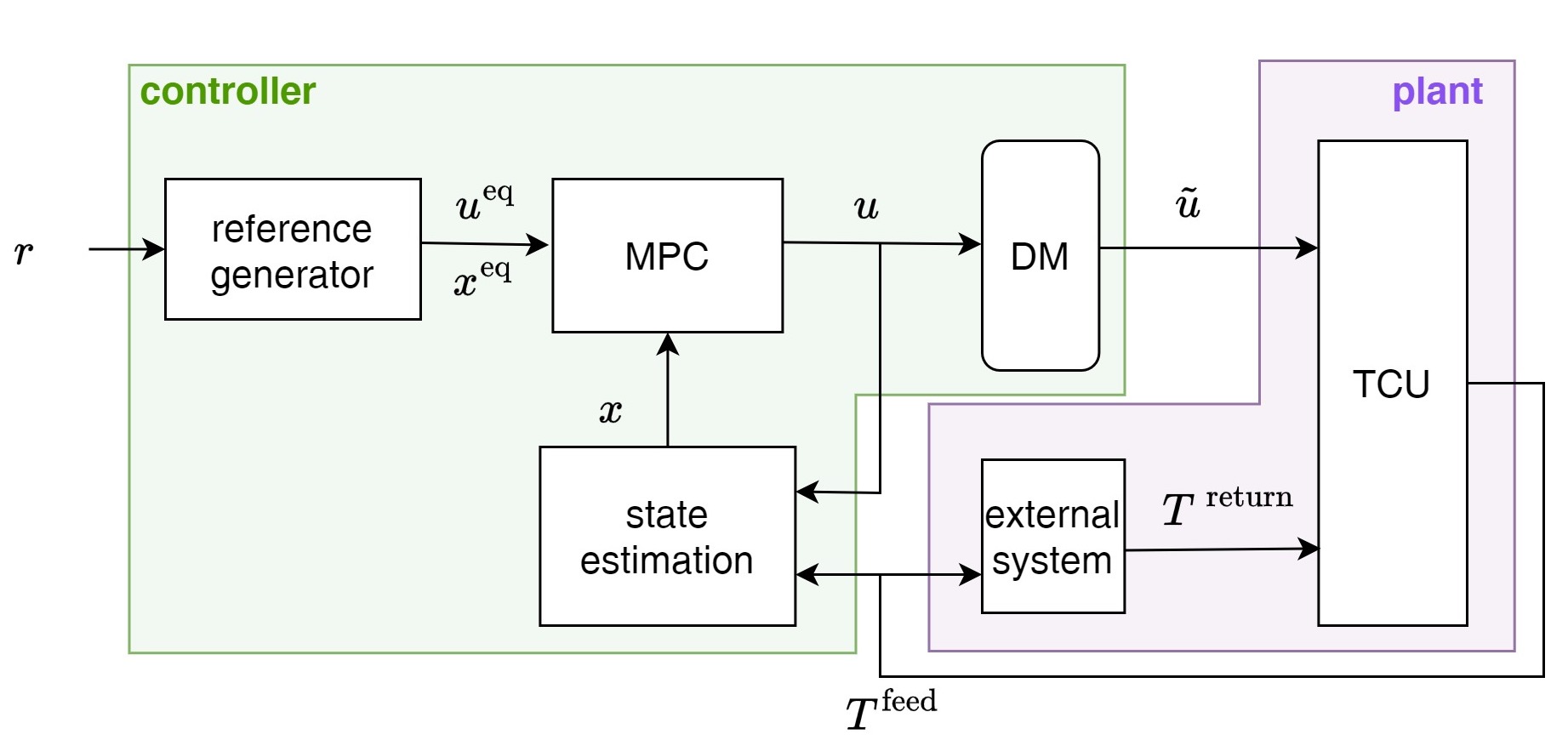}
	\caption{Schematic of the proposed control architecture}
	\label{fig:control_scheme}
\end{figure}

Now we design a Model Predictive Control (MPC) scheme to control our plant, based on the models identified in the previous section.
The schematic diagram of the control scheme is shown in Figure \ref{fig:control_scheme}. 

\subsection{Cost design}
The design of the MPC scheme should be driven by the minimization of a stage cost.
Our modeling choice for this is the sum of the squared tracking error with a linear input penalization, which represents an economic cost:
\begin{align}\label{eq:stage_cost}
    l( y, u, r) \coloneqq (y - r )^2 + \lambda \cdot (u^{\textup{heat}} + u^{\textup{cool}})
\end{align}
where $r$ is the reference signal, $u$ is the input of the system, $y$ is the output of the system and $\lambda$ is a parameter that we set to $\lambda = 50$.
This cost is chosen because the controller needs to not only track the temperature reference signal but also minimize energy consumption while performing this task.
\subsection{Reference generation for input and state}

For a certain output temperature reference $r$, we compute the equilibrium state that minimizes the stage cost \eqref{eq:stage_cost}. This is done by solving the following optimization problem
\begin{subequations} \label{eq:opti_equilibrium}
	\begin{align}
		\min_{u,x,y} & \quad  l(y, u, r)\label{eq:opt1:cost} \\
		\text{s.t.  }  &x =f(x,u)\label{eq:opt1:x}  \\
            &y = g(x)\label{eq:opt1:y} 
	\end{align}
\end{subequations}
The equilibrium state $x^{\textup{eq}}$ is used in the terminal constraint of the MPC optimization problem, which is defined in Section \ref{subsec:mpc}.

\subsection{State estimation}
In the case of the NNARX model, the state is directly given by past input and past measurements. In the case of the linear system, the state estimation is done via a Kalman Filter that takes into account the noise model that has been estimated (see \cite{simpson2023efficient} for more details).

\subsection{Model Predictive Control}\label{subsec:mpc}
An MPC control law is now designed.
It relies on solving
the following constrained optimization problem for each time step $k$
\begin{subequations} \label{eq:nomMPC}
	\begin{align}
 \underset{\substack{
              \bar{u}_0, \dots, \bar{u}_{N_p-1}, \\
              \bar{x}_0, \dots, \bar{x}_{N_p}
            }} {\text{minimize}}
            & \quad  \sum_{i=0}^{N_p-1 } 
            l\bigg( g(\bar{x}_i), \bar{u}_i, r  \bigg)
          \label{eq:nomMPC:cost} \\
		\text{subject to} & \\
		& \spaceOpti \bar{x}_{0} =x \label{eq:nomMPC:x0} \\
		& \spaceOpti \bar{x}_{i+1} = f(\bar{x}_{i}, \bar{u}_{i}) \quad
		\mbox{for} \; i = 1, \dots, N_p-1 \label{eq:nomMPC:dynamics} \\
		& \spaceOpti  \bar{u}_{i} \in \mathcal{U} \quad 
		\hspace{1.4cm} \mbox{for} \; i = 1, \dots, N_p-1 \label{eq:nomMPC:actuator} \\
		& \spaceOpti  \bar{x}_{N_p} = x^{\textup{eq}}\label{eq:nomMPC:terminalx}
	\end{align}
\end{subequations}
In the objective function \eqref{eq:nomMPC:cost}, $N_p=30$ denotes the prediction horizon.
The feasible input set $\mathcal{U} \coloneqq [0, 1]^2$ is chosen so that both $\bar{u}^{\textup{heat}}$ and $\bar{u}^{\textup{cool}}$ are constrained to be between $0$ and $1$.
The identified model is applied as a predictive model (see \eqref{eq:nomMPC:dynamics}) and it is initialized with the current state value $x$ (see \eqref{eq:nomMPC:x0}). Finally, \eqref{eq:nomMPC:terminalx} is introduced as terminal equality constraints to guarantee the nominal closed-loop stability (see \cite{rawlings2017model}).
According to the Receding Horizon Principle, only the first element $\bar{u}_{0}$ is applied to the plant. 
We write $\pi(\cdot, \cdot, \cdot )$ the function that inputs the current state, the current reference, and the corresponding equilibrium state and outputs the element $\bar{u}_0$ of the solution of \eqref{eq:nomMPC}:
\begin{align}
    \pi_{\mathrm{
MPC}}(x, r, x^{\textup{eq}}) \coloneqq \bar{u}_{0}.
\end{align}
Overall, the control policy induced by this MPC scheme reads as
\begin{align}
    u_k = \pi_{\mathrm{
MPC}}(x_k, r_k, x^{\textup{eq}}_k),
\end{align}
where $u_k$ is the input at time step $k$,
$x_k$ is the state of the system, $r_k$ is the reference, and $x^{\textup{eq}}_k$ is the desired equilibrium point computed via \eqref{eq:opti_equilibrium} with $r = r_k$.

Both the Linear MPC and the NN-MPC are implemented via the presented procedure. Only the system dynamic function $f(\cdot, \cdot)$ and the output function $g(\cdot)$ in \eqref{eq:opti_equilibrium} and \eqref{eq:nomMPC} differ from one controller to the other as they represent the two models discussed in Section \ref{sec:sysiden}.

Regarding the numerical optimization of problems \eqref{eq:opti_equilibrium} and \eqref{eq:nomMPC},
we used the nonlinear optimization solver IPOPT \cite{Waechter2006} together with the package CasADi \cite{Andersson2019} for the symbolic expressions and automatic differentiation, and the shipped sparse linear solver MUMPS.
As it is commonly done in MPC, the other components of the optimal solution of \eqref{eq:nomMPC} are also retrieved and used as a warm start for the next iteration to make the optimization routine converge faster.

\section{Experimental Results}\label{sec:exp}
\begin{figure}[t]
        \vspace{0.3cm}
	\centering	\includegraphics[width=0.9\columnwidth]{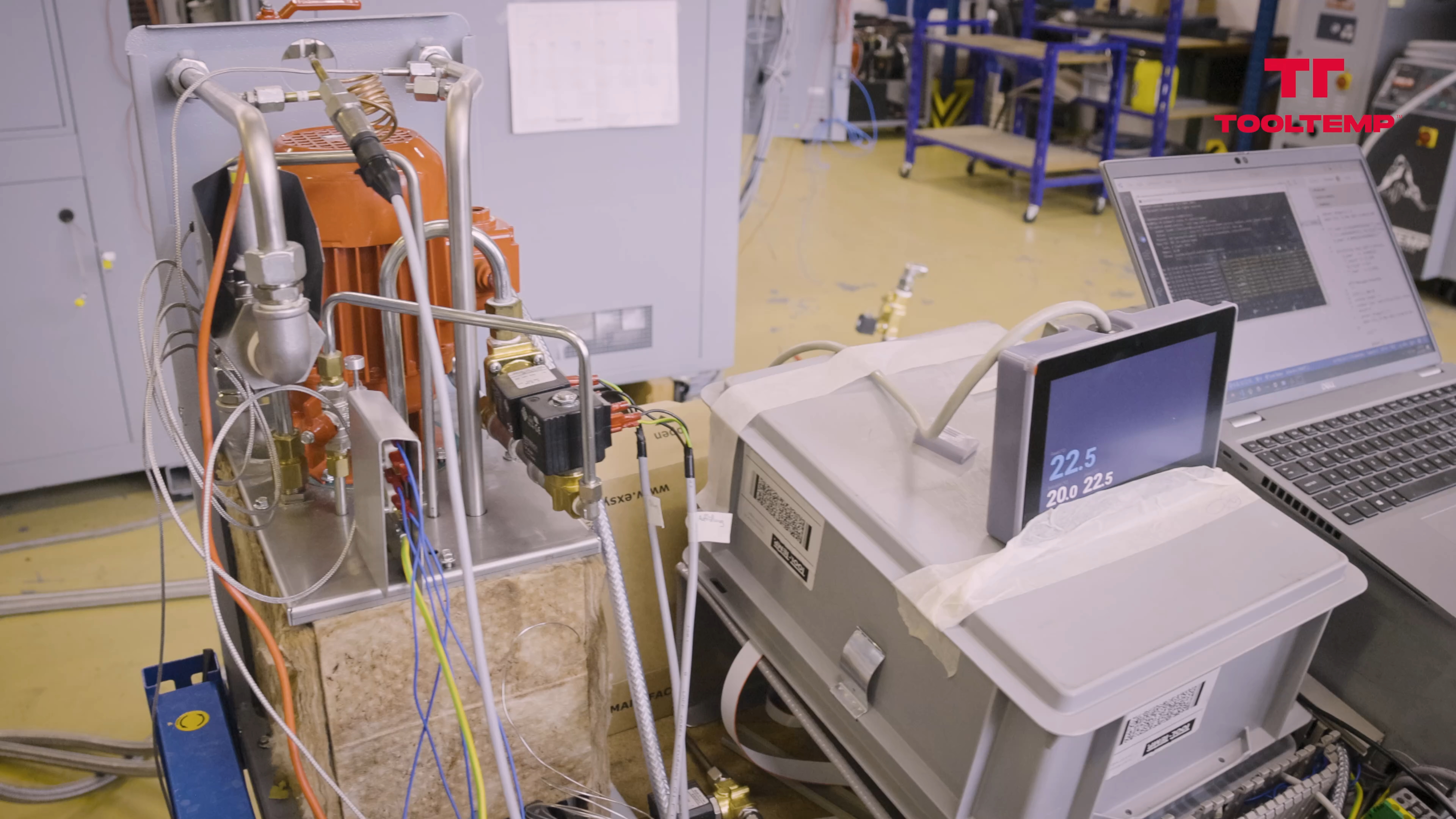}
	\caption{Experiment setup}
	\label{fig:tcu_experiment}
\end{figure}
\begin{figure}
	\begin{center}          \scalebox{\pgfscale}{\input{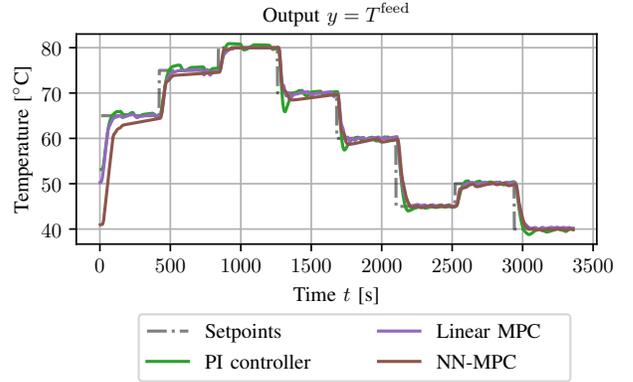}
            		} 
	\end{center}
        \vspace{-0.3cm}
	\caption{Temperature profiles for the three presented controllers.}\label{fig:controlled_trajectories}
 \vspace{-0.3cm}
\end{figure}
\begin{figure}
        \vspace{0.3cm}
	\begin{center}
             \scalebox{\pgfscale}{\input{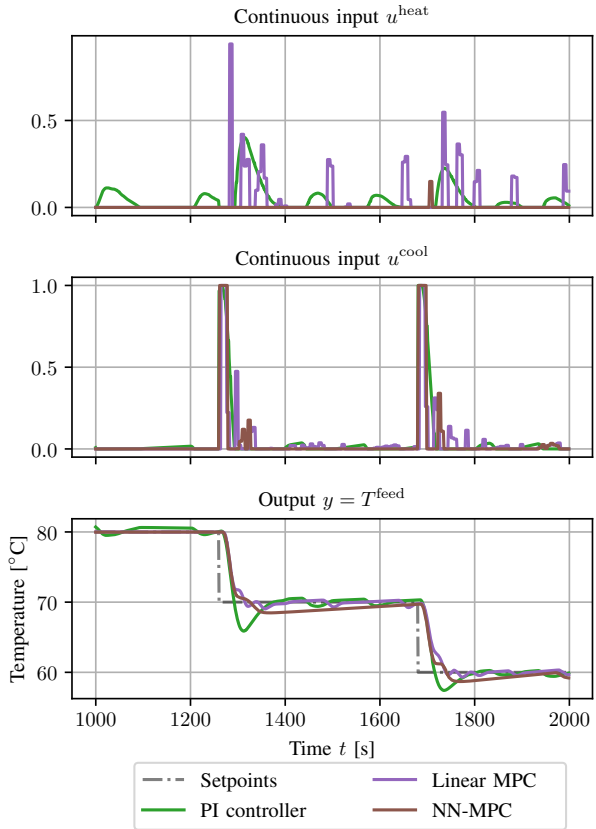}
            		}
	\end{center}
	\vspace{-0.3cm}
	\caption{Zoom-in on control performance}\label{fig:controlled_trajectories_zoom}
 \vspace{-0.3cm}
\end{figure}

To compare the presented controllers, we test them on the real system (see Figure \ref{fig:tcu_experiment}).
The same experiment is carried out for the PI controller defined in part \ref{subsec:PI} and for the MPC scheme with both linear and NNARX models.
The analog control input $u$ is computed via \eqref{eq:nomMPC} every $\tau_s = 6\textup{s}$ while the digital control input $\tilde{u}$ is computed every $\Delta t= 1\textup{s}$ according to \eqref{eq:modulation}.
When the solver still has not found a solution for \eqref{eq:nomMPC} after $3\textup{s}$, the input is left unchanged.
The results are depicted in Figures \ref{fig:controlled_trajectories} and Figure \ref{fig:controlled_trajectories_zoom}.

To assess the performance, we use the stage cost defined in \eqref{eq:stage_cost}, which is split in two parts: the tracking cost $(y - r )^2 $, and the energy cost $\lambda \cdot (u^{\textup{heat}} + u^{\textup{cool}})$.
These costs are averaged over time from $t_0 = 10$ minutes to $t_{\textup{end}} = 56$ minutes, the end of the experiment.
The choice of $t_0$ is made to discount the effects of the initial state of the system.
These results are presented in Table \ref{table:results}.
\begin{table}[t]
\vspace{0.4cm}
\centering
\begin{tabular}{||c c c c||} 
 \hline
 Controller & Tracking cost & Energy cost & Total cost \\ [0.5ex] 
 \hline\hline
 PI controller & $\bm{5.30}$ & $4.48$ & $9.78$ \\ 
 Linear MPC & $5.66$ & $3.93$ & $9.59$ \\
 NN-MPC & $5.97$ & $\bm{2.80}$ & $\bm{8.77}$ \\ [1ex] 
 \hline
\end{tabular}
\caption{Performance results of the presented controllers}
\label{table:results}
\vspace{-0.8cm}
\end{table}
The tracking performance of NN-MPC is worse compared to the other controllers. This could be attributed to increased computational effort, potentially leading to higher delays in implementing the control input.
However, the PI controller overshoots (or undershoots) more then MPC controllers. It is worth noting that NN-MPC uses less control effort, which results in reduced energy costs, thereby offering economic advantages in practical TCU operations.
Overall, the NN-MPC controller delivers the best performance.

\section{Conclusion}\label{sec:conclu}
In this paper, we introduce a novel NNARX-based model predictive control scheme designed specifically for Temperature Control Units produced by Tool-Temp AG. Our proposed scheme outperforms compared to both the traditional PI controller and linear MPC controller, for its ability to minimize energy costs. Despite its energy efficiency, the proposed scheme maintains a relatively decent reference tracking performance.

\section*{Acknowledgements}
\vspace{1mm}
\begin{minipage}[l]{0.075\textwidth}
	\includegraphics[width=\textwidth]{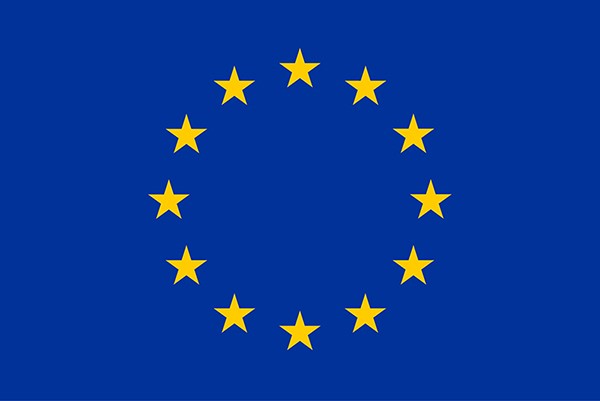}
	\label{fig:euflag}
\end{minipage}
\hspace{3mm}
\begin{minipage}[right]{0.34\textwidth}
	This project has received funding from the European Union’s Horizon 2020 research and innovation program under the Marie Skłodowska-Curie grant agreement No. 953348
\end{minipage}
\vspace{2mm}

\bibliographystyle{IEEEtran}
\bibliography{NN4TT}

\end{document}